\documentclass{ws-procs9x6-cpt22}
\begin{document}

\newcommand{\refeq}[1]{(\ref{#1})}
\def\etal {{\it et al.}}

\title{Gravity Tests with Radio Pulsars in Perturbative and Nonperturbative
Regimes}

\author{Lijing Shao$^{1,2}$}

\address{$^1$Kavli Institute for Astronomy and Astrophysics, Peking University,
Beijing 100871, China}

\address{$^2$National Astronomical Observatories, Chinese Academy of Sciences,
Beijing 100012, China}

\begin{abstract}
Searches for empirical clues beyond Einstein's general relativity (GR) are
crucial to understand gravitation and spacetime.  Radio pulsars have been
playing an important role in testing gravity theories since 1970s. Because radio
timing of binary pulsars is very sensitive to changes in the orbital dynamics,
small deviations from what GR predicts can be captured or constrained.  In this
sense, the gravity sector in the standard-model extension was constrained
tightly with a set of pulsar systems.  Moreover, compact objects like pulsars
are possible to develop nonperturbative deviations from GR in some specific
alternative gravity theories, thus radio pulsars also provide rather unique
testbeds in the strong-gravity regime.
\end{abstract}

\bodymatter

\section{Introduction}

Among the four fundamental forces in the Nature, gravity is rather unique as it
is described in the language of differential geometry, while the other three
forces are understood in terms of quantum field theory. Therefore, to go beyond
the current paradigm of modern physics, which consists of Einstein's general
relativity (GR) and the standard model of particle physics, gravity might hold
the key.  Empirical studies of gravitation and spacetime are important to
provide clues to a deep fundamental theory, probably the quantum
gravity.\cite{review,datatables} In testing gravity theories, radio pulsars have
been playing an important and unique role since the discovery of the
Hulse-Taylor pulsar in 1970s. In this short proceedings, we will briefly review
some interesting bounds from pulsar observations in a perturbative framework,
called the standard-model extension (SME),\cite{sme} as well as in some specific
scalar-tensor gravity theories where nonperturbative strong-field phenomena
might develop inside neutron stars. Pulsar timing puts remarkable limits in both
perturbative and nonperturbative gravity regimes.

\section{Perturbative weak-field expansion of gravity}

As GR has been confronted with various kinds of experiments and observations for
a century where all tests are passed with flying colors,\cite{review,datatables}
one might only expect small deviations from it, at least in the weak-field
limit.  The gravity sector of SME is designed in the spirit of effective field
theory, and it categorizes all kinds of operators beyond GR by introducing SME
coefficients for Lorentz/CPT violation.\cite{sme} In the pure gravity sector,
the most generic Lagrangian for linearized gravity reads,
\begin{equation}
	\label{eq:sme}
	\mathcal{L}_{\mathcal{K}^{(d)}}=\frac{1}{4} h_{\mu \nu}
	{\hat{\mathcal{K}}}^{(d) \mu \nu \rho \sigma} h_{\rho \sigma} \,,
\end{equation}
where $\hat{\mathcal{K}}^{(d) \mu \nu \rho \sigma}=\mathcal{K}^{(d) \mu \nu \rho
\sigma i_{1} i_{2} \cdots i_{d-2}} \partial_{i_{1}} \partial_{i_{2}} \cdots
\partial_{i_{d-2}}$ is a complicated operator with derivatives contracted with
SME coefficients $\mathcal{K}^{(d) \mu \nu \rho \sigma i_{1} i_{2} \cdots
i_{d-2}}$. The complete action~(\ref{eq:sme}) can be very cumbersome and
contains an infinite number of field operators. However, in the sense of
effective field theory, it is likely that terms of the lowest mass dimensions
dominate in certain low-energy experiments.

In a modified gravity, a binary orbit is generally altered. This results in
characteristic changes in the times of arrival, the main observables, for binary
pulsars. In turn, dedicated long-term observations of radio pulsars can provide
stringent limits to various types of modifications in the gravitational
interaction. An updated list of gravity tests in the SME framework provided by
pulsars includes tests of,\cite{pulsarsme}
\begin{itemize}
	\item the minimal gravity sector with  operators of mass dimension four, 
	\item CPT-violating operators of mass dimension five,
	\item nonlinear operators of mass dimension eight which violate the
	gravitational weak equivalence principle,
	\item matter-gravity couplings with operators of mass dimensions three and
	four, and
	\item abnormal spin behaviours caused by the Lorentz-violating neutron star
	structure, or due to gravity and matter-gravity couplings.  
\end{itemize}
A summary of limits from pulsar timing experiments can be found in the {\it Data
Tables for Lorentz and CPT Violation},\cite{datatables} and for details readers
are referred to original publication.

\section{Nonperturbative strong-field gravity}

\begin{figure}
	\centering
	\includegraphics[width=8cm]{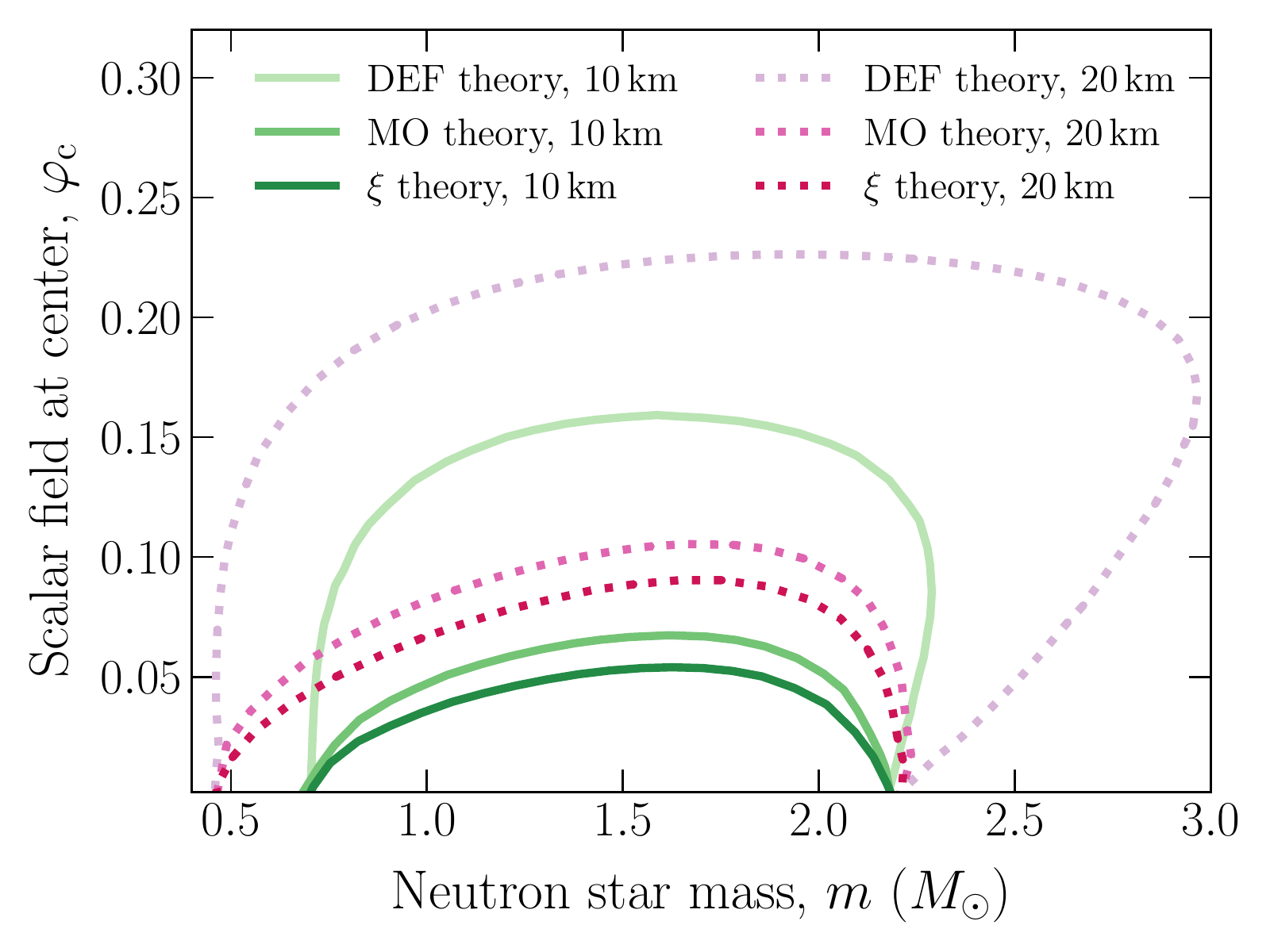}
	\caption{Scalar field at the center of neutron stars in three massive
	scalar-tensor theories with a reduced Compton wavelength of $10\,$km ({\it
	solid green lines}) and $20\,$km ({\it dotted red lines}).\cite{massiveSTG}
	\label{fig:massiveSTG}}
\end{figure}

The treatment in the SME has assumed the smallness of any kinds of deviations
from GR. However, neutron stars are strongly self-gravitating objects. As
discovered by Damour and Esposito-Far\`ese in 1990s, a nonperturbative phenomena
called ``spontaneous scalarization'' might happen for neutron stars in a class
of scalar-tensor gravity theories.\cite{def} This behaviour introduces an extra
dipolar channel for gravitational radiation in a binary and can be constrained
by pulsar timing, via the orbital decay rate parameter, $\dot P_{\rm b}$.  There
are a few variants of scalar-tensor gravity theories, including those with a
massive scalar field\cite{massiveSTG} and with a topological Gauss-Bonnet
term.\cite{gb} Scalarized neutron stars are illustrated in
Fig.~\ref{fig:massiveSTG} for three representative massive scalar-tensor
theories, including Damour-Esposito-Far\`ese theory, Mendes-Ortiz theory, and a
$\xi$ theory from considerations in cosmology.  As we can see, scalar hairs grow
for neutron stars with certain masses as they are energetically favored. Current
pulsar-timing observations of a handful of neutron-star white-dwarf binaries and
asymmetric double neutron star binaries are able to put stringent constraints on
theory parameters.\cite{def, gb} Recently, gravitational waves also start to
provide useful limits,\cite{gw} and in many cases, depending on the specifics of
theories under investigation, limits from pulsar timing and gravitational waves
are complementary to each other.

\section{Discussion}

Neutron stars are superb testbeds for gravitation and spacetime.  Thanks to the
precision timing ability of large-area radio telescopes, gravity tests are
versatile with radio pulsars. A number of changes in the orbital dynamics of
different types can be probed. In particular, it was demonstrated for a couple
of times that, a set of carefully chosen binary pulsars are able to break
degeneracy of theory parameters and  put combined limits on the SME coefficients
for Lorentz/CPT violations. These limits usually are very tight and provide
important experimental results for the SME community. On the other hand, in some
specific alternative theories of gravity, the perturbative treatment fails, and
nonperturbative hairs grow for certain neutron stars. In such a case, pulsar
timing appears even advantageous for empirical gravity tests, and provides
remarkable constraints for gravity in the strong-field regimes, complementing
the new tests brought by observations of gravitational waves and black hole
shadows. 

In a short summary, both perturbative and nonperturbative probes of the
gravitational interaction are useful and might lead to clues for quantum
gravity. Radio pulsars, whose timing results are extremely precise and improve
over time, stand as a unique testbed for gravity. In the upcoming years, we can
certainly expect improved tests from existing pulsar systems, as well as new
tests from yet-to-be-discovered pulsars, for example, possibly from pulsars in
binary with black holes.

\section*{Acknowledgments}

I am grateful to Quentin Bailey, Alan Kosteleck\'y, Norbert Wex for stimulating
discussions in the past few years.  This work was supported by the National SKA
Program of China (2020SKA0120300), the National Natural Science Foundation of
China (11975027, 11991053, 11721303), and the Max Planck Partner Group Program
funded by the Max Planck Society.


\begin{thebibliography}{xx}

\bibitem{review}
C.~M.~Will,
Living Rev. Rel. \textbf{17}, 4 (2014);
E.~Berti, \etal, 
Class. Quant. Grav. \textbf{32}, 243001 (2015).

\bibitem{datatables}
{\it Data Tables for Lorentz and CPT Violation,}
V.A.\ Kosteleck\'y and N.\ Russell,
2022 edition,
arXiv:0801.0287v15.

\bibitem{sme}
D.~Colladay and V.~A.~Kosteleck\'y,
Phys. Rev. D \textbf{55}, 6760 (1997);
Phys. Rev. D \textbf{58}, 116002 (1998);
V.~A.~Kosteleck\'y,
Phys. Rev. D \textbf{69}, 105009 (2004);
Q.~G.~Bailey and V.~A.~Kosteleck\'y,
Phys. Rev. D \textbf{74}, 045001 (2006);
V.~A.~Kosteleck\'y and M.~Mewes,
Phys. Lett. B \textbf{779}, 136 (2018);
V.~A.~Kosteleck\'y and Z.~Li,
Phys. Rev. D \textbf{99},  056016 (2019);
Phys. Rev. D \textbf{104}, 044054 (2021).

\bibitem{pulsarsme}
B.~Altschul,
Phys. Rev. D \textbf{75}, 023001 (2007);
L.~Shao,
Phys. Rev. Lett. \textbf{112}, 111103 (2014);
Phys. Rev. D \textbf{90}, 122009 (2014);
L.~Shao and Q.~G.~Bailey,
Phys. Rev. D \textbf{98}, 084049 (2018);
Phys. Rev. D \textbf{99}, 084017 (2019);
L.~Shao,
Symmetry \textbf{11},  1098 (2019);
R.~Xu, \etal,
Phys. Lett. B \textbf{803}, 135283 (2020).

\bibitem{def}
T.~Damour and G.~Esposito-Far\`ese,
Phys. Rev. Lett. \textbf{70}, 2220 (1993);
Phys. Rev. D \textbf{54}, 1474 (1996);
P.~C.~C.~Freire, \etal.,
Mon. Not. Roy. Astron. Soc. \textbf{423}, 3328 (2012);
N.~Wex,
arXiv:1402.5594 [gr-qc];
L.~Shao, \etal,
Phys. Rev. X \textbf{7}, 041025 (2017);
M.~Kramer, \etal,
Phys. Rev. X \textbf{11}, 041050 (2021);
J.~Zhao, \etal,
Class. Quant. Grav. \textbf{39}, 11LT01 (2022).

\bibitem{massiveSTG}
F.~M.~Ramazano\u{g}lu and F.~Pretorius,
Phys. Rev. D \textbf{93}, 064005 (2016);
S.~S.~Yazadjiev, \etal,
Phys. Rev. D \textbf{93},  084038 (2016);
R.~F.~P.~Mendes and N.~Ortiz,
Phys. Rev. D \textbf{93}, 124035 (2016);
R.~Xu, \etal,
Phys. Rev. D \textbf{102}, 064057 (2020);
Z.~Hu, \etal,
Phys. Rev. D \textbf{104}, 104014 (2021).

\bibitem{gb}
D.~D.~Doneva and S.~S.~Yazadjiev,
Phys. Rev. Lett. \textbf{120}, 131103 (2018);
H.~O.~Silva, \etal,
Phys. Rev. Lett. \textbf{120}, 131104 (2018);
R.~Xu, \etal,
Phys. Rev. D \textbf{105}, 024003 (2022);
V.~I.~Danchev, \etal,
arXiv:2112.03869 [gr-qc].

\bibitem{gw}
C.~M.~Will,
Phys. Rev. D \textbf{50}, 6058 (1994);
T.~Damour and G.~Esposito-Far\`ese,
Phys. Rev. D \textbf{58}, 042001 (1998);
S.~Tahura and K.~Yagi,
Phys. Rev. D \textbf{98}, 084042 (2018);
B.~P.~Abbott \etal,
Phys. Rev. Lett. \textbf{123}, 011102 (2019);
C.~Liu, \etal,
Mon. Not. Roy. Astron. Soc. \textbf{496}, 182 (2020);
S.~E.~Perkins, \etal,
Phys. Rev. D \textbf{103}, 044024 (2021);
J.~Zhao, \etal,
Phys. Rev. D \textbf{104}, 084008 (2021);
R.~Abbott \etal,
arXiv:2112.06861 [gr-qc].
	
\end{thebibliography}
\end{document}